# Quantum Theory of Half-integer Spin Free Particles from the Perspective of the Majorana Equation

*Luca Nanni
Faculty of Natural Science, University of Ferrara, 44100 – Ferrara, Italy

*corresponding author e-mail: luca.nanni@student.unife.it

**Abstract**

In this study, the Majorana equation for particles with arbitrary spin is solved for a half-integer spin free particle. The solution for the fundamental state, corresponding to the reference frame in which the particle is at rest, is compared with that obtained using the Dirac equation, especially as regards the approximation in the relativistic limit, in which the speed of the particle is close to that of light. Furthermore, the solutions that Majorana defines unphysical, proving that their occupation probability increases with the particle velocity, are taken into consideration. The anomalous behavior exhibited by these states suggests that for high-energy particles with small mass, transitions from a bradyonic state to a tachyonic state could become possible.

## 1 Introduction

In 1932, Majorana formulated an equation for particles with arbitrary spin. This equation is relativistically invariant and valid for both bosons and fermions [1-2]. However, if the spin and mass of a particle are set, this equation leads to an infinite set of solutions with a spectrum of decreasing masses. In his article, Majorana considers these solutions unphysical or accessible only under extreme conditions, focusing the attention on the fundamental state, where the reference system is that of the particle at rest. Furthermore, the equation allows solutions with imaginary mass, compatible with the tachyonic behavior of matter. These innovative results have not been properly considered by the physicists of the 1930s and the work of Majorana has received little attention for a least a decade [2–4]. Moreover, the Dirac equation was proved effective in explaining the spectrum of the hydrogen atom correctly. Additionally, the existence of the positron, whose discovery was announced in 1932, was theorized using the Dirac equation. However, the Majorana theory was recently reconsidered, particularly in the context of high-spin particle physics [4–17].

In this paper, the relativistic quantum theory of Majorana for a free massive particle with half-integer spin is presented. The results, where possible, are compared with those obtained by solving the relativistic Dirac equation [18]. If the spin and mass of the particle are set, the Majorana equation provides the solution for both the fundamental state and infinite series of states considered unphysical by Majorana and referred to as excited states with increasing angular quantum number $J$ by us. Each solution corresponds to a given projection of $J$ along the z-axis, and is represented by a vector with infinite components. The nonzero components of this vector correspond to the nonzero elements of infinite matrices $\boldsymbol{\alpha_k}$ and $\boldsymbol{\beta}$. The method of calculation of these vectors is explained in the following sections. The Majorana equation is relativistically invariant, used for free particles, and valid for any value of the spin and velocity of the particles. This equation does not allow solutions with negative energy (Majorana antiparticle). Furthermore, unlike the Dirac equation, the Majorana equation was formulated without the need of compliance with the formula of the relativistic energy: $E^2 = p^2c^2 + m^2c^4$.

For each solution corresponding to the same $m_J$ (magnetic quantum number), the wavevector given by the linear combination of all infinite solutions, in compliance with the principle of superposition of states, is constructed. Then, the probability of occupation of each possible state as a function of the Lorentz factor $(v/c)$ and the energies of transitions between any possible states are calculated. Moreover, the physical behavior of the particle in the limit $v \rightarrow c$ is analyzed in detail using the Heisenberg uncertainty principle.

## 2 Solution of the Majorana Equation for the Fundamental State

The explicit form of Majorana equation with infinite components is [1-2]:

$$\left(\mathbb{1} i\hbar\frac{\partial}{\partial t} - c\boldsymbol{\alpha}_{\mathbf{1}} i\hbar\frac{\partial}{\partial x} - c\boldsymbol{\alpha}_{\mathbf{2}} i\hbar\frac{\partial}{\partial y} - c\boldsymbol{\alpha}_{\mathbf{3}} i\hbar\frac{\partial}{\partial z} - \boldsymbol{\beta} m_0 c^2\right)|\Psi\rangle = 0 \qquad \textbf{(1)}$$

where $\boldsymbol{\alpha}_{\boldsymbol{k}}$ and $\boldsymbol{\beta}$ are infinite matrices, $\mathbb{1}$ is the infinite unity matrix and $|\Psi\rangle$ is the infinite components spinor that must satisfy unitary condition. In particular, we want to study the fundamental state of a free particle with half-integer spin that is consistent with that reported in the original Majorana article and corresponds to the case in which the reference system is that of the center of mass. This constraint implies that the nonzero components of matrices $\boldsymbol{\alpha}_{\boldsymbol{k}}$ must coincide with those of Dirac. Matrix $\boldsymbol{\beta}$ must be definite and positive (all positive eigenvalues), and must satisfy the algebraic relationship $\boldsymbol{\beta}^{-1} = \boldsymbol{\beta}^{\dagger}$. Since these matrices are infinite, the Majorana equation, whose form is the same as the form of the Dirac equation, is reduced to an infinite system of linear differential equations. However, because the nonzero components of the matrices are related to the first four rows and four columns, the system of infinite equations is reduced to a system similar to that of Dirac, with the difference in the structure of matrix $\boldsymbol{\beta}$. Consequently, also spinor $|\Psi\rangle$ reduces to a four-non-trivial components vector since all the infinite components, except the first four, are multiplied times the zero components of the $\boldsymbol{\alpha}_{\boldsymbol{k}}$ and $\boldsymbol{\beta}$ matrices. This implies the wavefunction unitary violation that, however, may be recovered normalizing the *new* reduced ket $|\Psi\rangle$, formed by four-non-trivial components, as done for the Dirac wavefunction. The easiest and most direct way to assure positive values of the energy, as required by the Majorana theory, is to force infinite matrix $\boldsymbol{\beta}$ to be the unit matrix:

$$\boldsymbol{\beta} = \begin{pmatrix} 1 & 0 & 0 & \cdots \\ 0 & 1 & 0 & \cdots \\ 0 & 0 & 1 & \cdots \\ \vdots & \vdots & \vdots & \ddots \end{pmatrix}$$

This form is  consistent with the form predicted by the Majorana theory [1]:

$$\boldsymbol{\beta} = \frac{\mathbb{1}}{s + \frac{1}{2}}$$

where $s$ is the particle spin. The explicit form of matrices $\boldsymbol{\alpha_k}$is given by:

$$\boldsymbol{\alpha_k} = \begin{pmatrix} \mathbf{0} & \boldsymbol{\sigma_k} & \cdots \\ \boldsymbol{\sigma_k} & \mathbf{0} & \cdots \\ \vdots & \vdots & \ddots \end{pmatrix}$$

where $\boldsymbol{\sigma_k}$ are the Pauli matrices [18]. The 4-vector signature is $(+,-,-,-)$. By inserting these infinite matrices into Eq. **(1)**, we obtain a system of four linear differential equations of the first order. Considering only the nonzero elements of the system, the associated Hamiltonian matrix is given by:

$$\boldsymbol{H} = \hat{E} = \begin{pmatrix} m_0c^2 & 0 & c\hat{p}_z & c(\hat{p}_x + i\hat{p}_y) \\ 0 & m_0c^2 & c(\hat{p}_x - i\hat{p}_y) & -c\hat{p}_z \\ c\hat{p}_z & -c(\hat{p}_x + i\hat{p}_y) & m_0c^2 & 0 \\ -c(\hat{p}_x - i\hat{p}_y) & -c\hat{p}_z & 0 & m_0c^2 \end{pmatrix} \tag{2}$$

where operator $\hat{E}$ is represented by the matrix:

$$\hat{E} = \begin{pmatrix} E & 0 & 0 & 0 \\ 0 & E & 0 & 0 \\ 0 & 0 & E & 0 \\ 0 & 0 & 0 & E \end{pmatrix}$$

The solutions of Eq. **(1)** are plane waves [19]:

$$\Psi_j = u_j(\boldsymbol{p}) exp\left[\mp i \frac{\frac{E}{c}t - \boldsymbol{x}\cdot\boldsymbol{p}}{\hbar}\right] \tag{3}$$

where $\boldsymbol{x}$ and $\boldsymbol{p}$ are the space-like components of the 4-position and of the 4-momentum, respectively. Using the following expressions:

$$(\hat{E} + m_0c^2)|\Psi\rangle = (E + m_0c^2)|\Psi\rangle$$

$$\hat{p}_k|\Psi\rangle = p_k|\Psi\rangle$$

the four solutions of the Majorana equation, whose spinors (limited to the first four non-trivial components) are obtained as:

$$|u\rangle_{up}^{M} = \sqrt{E+m_0c^2}\begin{pmatrix} 1 \\ 0 \\ -\frac{cp_z}{E+m_0c^2} \\ \frac{c(p_x-ip_y)}{E+m_0c^2} \end{pmatrix} \quad , \quad |u\rangle_{down}^{M} = \sqrt{E+m_0c^2}\begin{pmatrix} 0 \\ 1 \\ \frac{c(p_x+ip_y)}{E+m_0c^2} \\ \frac{cp_z}{E+m_0c^2} \end{pmatrix} \qquad \textbf{(4)}$$

$$|u\rangle_{up\prime}^{M} = \sqrt{E+m_0c^2}\begin{pmatrix} -\frac{cp_z}{m_0c^2+E} \\ -\frac{c(p_x-ip_y)}{m_0c^2+E} \\ 1 \\ 0 \end{pmatrix} \quad , \quad |u\rangle_{down\prime}^{M} = \sqrt{E+m_0c^2}\begin{pmatrix} -\frac{c(p_x+ip_y)}{m_0c^2+E} \\ \frac{cp_z}{m_0c^2+E} \\ 0 \\ 1 \end{pmatrix} \qquad \textbf{(5)}$$

where $\sqrt{E+m_0c^2}$ is the normalization constant proportional to the particle energy so the total probability density is Lorentz invariant [17] (under the approximation to consider only the first four components). Since the energy is always positive, to distinguish the particle and antiparticle states, the *apostrophe* on the antiparticle spinors with the opposite spin is introduced. Subscripts *up* and *down* indicate the two possible spin states. Moreover, to distinguish the Majorana spinors from the Dirac spinors unambiguously, *M* is used to indicate the Majorana wavevector. Similar to the solutions of the Dirac equation [9], those of the Majorana equation are reduced to the Schrödinger eigenfunctions when the velocity of the particle is much lower than that of light:

$$\frac{cp_k}{m_0c^2 \pm E} = c\frac{m_0v_k}{m_0c^2 \pm E} = \frac{m_0v_k}{m_0c \pm \frac{E}{c}} \cong 0 \qquad v_k \ll c$$

Let us consider the case when $v_k \cong c$. Using the approximation:

$$\frac{cp_k}{mc^2 \pm E} = \frac{\gamma m_0 c v_k}{m_0c^2 \pm \gamma m_0 c^2} \cong \pm\frac{v_k}{c} \qquad v_k \cong c$$

where $\gamma$ is the Lorentz factor, we obtain the following spinors:

$$|u\rangle_{up}^{M} = \frac{1}{2}\begin{pmatrix} 1 \\ 0 \\ -1 \\ 1-i \end{pmatrix} \quad , \quad |u\rangle_{down}^{M} = \frac{1}{2}\begin{pmatrix} 0 \\ 1 \\ 1+i \\ 1 \end{pmatrix} \qquad v_k \cong c \tag{6}$$

$$|u\rangle_{up'}^{M} = \frac{1}{2}\begin{pmatrix} -1 \\ -(1-i) \\ 1 \\ 0 \end{pmatrix} \quad , \quad |u\rangle_{down'}^{M} = \frac{1}{2}\begin{pmatrix} -(1+i) \\ 1 \\ 0 \\ 1 \end{pmatrix} \qquad v_k \cong c \tag{7}$$

The phase of each spinor is the same for the solutions of the Dirac equation, i.e., $e^{\mp i\boldsymbol{p}\cdot\boldsymbol{x}/\hbar}$.

Let us compare the Majorana spinors with the Dirac spinors. The latter are given by:

$$|u\rangle_{up(+)} = \sqrt{E + m_0c^2}\begin{pmatrix} 1 \\ 0 \\ -\frac{cp_z}{E+m_0c^2} \\ \frac{c(p_x-ip_y)}{E+m_0c^2} \end{pmatrix} \quad , \quad |u\rangle_{down(+)} = \sqrt{E + m_0c^2}\begin{pmatrix} 0 \\ 1 \\ \frac{c(p_x+ip_y)}{E+m_0c^2} \\ \frac{cp_z}{E+m_0c^2} \end{pmatrix} \tag{8}$$

$$|u\rangle_{up(-)} = \sqrt{-E + m_0c^2}\begin{pmatrix} -\frac{cp_z}{m_0c^2-E} \\ -\frac{c(p_x-ip_y)}{m_0c^2-E} \\ 1 \\ 0 \end{pmatrix} , |u\rangle_{down(-)} = \sqrt{-E + m_0c^2}\begin{pmatrix} -\frac{c(p_x+ip_y)}{m_0c^2-E} \\ \frac{cp_z}{m_0c^2-E} \\ 0 \\ 1 \end{pmatrix} \tag{9}$$

Subscripts (+) and (–) indicate the positive and negative values of the energy, respectively. The first two expressions exhibit the states of the particle with spin up and spin down, while the latter exhibit the states of the antiparticle with spin up and spin down. In the limit $v_k \cong c$, the Dirac spinors become:

$$|u\rangle_{up(+)} = \frac{1}{2}\begin{pmatrix} 1 \\ 0 \\ -1 \\ 1-i \end{pmatrix} \quad , \quad |u\rangle_{down(+)} = \frac{1}{2}\begin{pmatrix} 0 \\ 1 \\ 1+i \\ 1 \end{pmatrix} \qquad v_k \cong c$$

$$|u\rangle_{up(-)} = \frac{1}{2}\begin{pmatrix} 1 \\ 1-i \\ 1 \\ 0 \end{pmatrix} \quad , \quad |u\rangle_{down(-)} = \frac{1}{2}\begin{pmatrix} 1+i \\ -1 \\ 0 \\ 1 \end{pmatrix} \qquad v_k \cong c$$

Although the eigenfunctions of a free particle are not normalizable, it is easy to verify that the scalar $\langle u|u\rangle$ concerning states of the antiparticle, which is proportional to the probability that

state $|u\rangle$ is occupied, changes depending on the considered spinor. For instance, for state *up* of the Majorana antiparticle we have:

$$\langle u|u\rangle_{up}^{M} = \frac{c^2p^2 + (m_0c^2 + E)^2}{(m_0c^2 + E)}$$

while in the Dirac theory, such a scalar becomes:

$$\langle u|u\rangle_{up(-)} = \frac{c^2p^2 + (m_0c^2 - E)^2}{(m_0c^2 - E)}$$

The difference between the two scalars is:

$$\Delta\langle u|u\rangle_{up}^{D-M} = \frac{4c^2p^2m_0c^2E}{(m_0^2c^4 - E^2)}$$

This difference is always positive, which means that the probability that an antiparticle state is occupied is always smaller in the case of the Majorana theory. For particle states, the predicted probabilities are the same. In the limit $v_k \cong c$, the probabilities of occupation of antiparticle states are equal for both theories, though the energies have opposite signs.

## 3 Majorana Excited States

The Majorana theory for particles with arbitrary spin predicts the existence of excited states with total angular momentum $J_n = s + n$ $(n = 1,2,\dots)$, to which the values of the mass correspond [1]:

$$m_{J_n} = \frac{m_0}{J_n + \frac{1}{2}} \qquad \textbf{(10)}$$

Equation **(10)** shows that for the fundamental state, mass $m_s$ equals the mass of the particle at rest, and it decreases progressively for the excited states:

$$m_{J_n} = \frac{m_0}{n+1} \qquad \textbf{(11)}$$

Equation **(11)** does not replace the relativistic formula $m = \gamma m_0$, where $\gamma$ is the Lorentz factor, which describes the increase of the inertial mass of the particle with the speed. Let us analyze in detail the first excited state with $J = 3/2$. Infinite matrices $\boldsymbol{\sigma_k}$ corresponding to this state are easily obtained from the Majorana relationships [1-2], setting $J = 3/2$ and $m_J = -\frac{3}{2}, -\frac{1}{2}, \frac{1}{2}, \frac{3}{2}$:

$$\boldsymbol{\sigma_1}(3/2) = \frac{1}{2}\begin{pmatrix} 0 & \sqrt{3} & 0 & 0 \\ \sqrt{3} & 0 & 2 & 0 \\ 0 & 2 & 0 & \sqrt{3} \\ 0 & 0 & \sqrt{3} & 0 \end{pmatrix} \quad ; \quad \boldsymbol{\sigma_2}(3/2) = \frac{i}{2}\begin{pmatrix} 0 & -\sqrt{3} & 0 & 0 \\ \sqrt{3} & 0 & -2 & 0 \\ 0 & 2 & 0 & -\sqrt{3} \\ 0 & 0 & \sqrt{3} & 0 \end{pmatrix}$$

$$\boldsymbol{\sigma_3}(3/2) = \frac{1}{2}\begin{pmatrix} 3 & 0 & 0 & 0 \\ 0 & 1 & 0 & 0 \\ 0 & 0 & -1 & 0 \\ 0 & 0 & 0 & -3 \end{pmatrix}$$

For convenience, we omitted all other infinite zero elements of the matrices. Using these matrices, $\boldsymbol{\alpha_k}$ are constructed:

$$\boldsymbol{\alpha_k} = \begin{pmatrix} \mathbf{0} & \boldsymbol{\sigma_k} \\ \boldsymbol{\sigma_k} & \mathbf{0} \end{pmatrix}$$

The explicit form of matrix $\boldsymbol{\beta}$ is [1]:

$$\boldsymbol{\beta}(3/2) = \frac{\mathbb{1}}{\frac{3}{2} + \frac{1}{2}} = \frac{1}{2}\mathbb{1}$$

whose nonzero elements are the first eight elements on the main diagonal. By inserting these matrices into the Majorana equation, we obtain an infinite system of linear differential equations, whose first eight terms are nonzero and provide nontrivial solutions. Considering only these terms, the eight spinors can be written as (the normalization constant has been omitted):

$$\begin{cases} |u(3/2)\rangle_{up}^{M} = (1,0,0,0,B_1,B_2,B_3,B_4)^T \\ |u(1/2)\rangle_{up}^{M} = (0,1,0,0,B_2,B_1,B_4,B_3)^T \\ |u(-1/2)\rangle_{down}^{M} = (0,0,1,0,B_3^*,-B_4^*,B_1^*,-B_2^*)^T \\ |u(-3/2)\rangle_{down}^{M} = (0,0,0,1,B_4^*,-B_3^*,B_2^*,-B_1^*)^T \end{cases} \qquad \mathbf{(12)}$$

$$\begin{cases} |u(3/2)\rangle_{up'}^{M} = (B_1,-B_2,B_3,-B_4 1,0,0,0)^T \\ |u(1/2)\rangle_{up'}^{M} = (B_2,-B_1,B_4,-B_3,0,1,0,0)^T \\ |u(-1/2)\rangle_{down'}^{M} = (-B_3^*,-B_4^*,-B_1^*,-B_2^*,0,0,1,0)^T \\ |u(-3/2)\rangle_{down'}^{M} = (-B_4^*,-B_3^*,-B_2^*,-B_1^*,0,0,0,1,)^T \end{cases} \qquad \mathbf{(13)}$$

Complex values $B_1, B_2, B_3$, and $B_4$ are the solutions of the system of linear equations and, similar to the solutions obtained in the previous section, they are quadratic functions of the individual components of the energy and of the linear momentum:

$$B_1, B_2 \propto \left(\frac{cp_z}{E + m_0c^2}\right)^2 \quad \overset{v\to c}{\Longrightarrow} \quad \left(\frac{v_z}{c}\right)^2 \cong 1$$

$$B_3, B_4 \propto \left(\frac{c\left(p_x + ip_y\right)}{E + m_0c^2}\right)^2 \quad \overset{v\to c}{\Longrightarrow} \quad (1 + i)^2 \cong 2i$$

Equations **(12)** represent particle states, while Eqs. **(13)** represent antiparticle states. Thus, assuming that the possible quantum states of the particle motion are the fundamental state and the first excited state, the spinor with infinite components in the limit $v \to c$ is:

$$|u(1/2,3/2)\rangle_{up}^{M} = (1,0,1,1 + i, 1,0,0,0,1,1, (1 + i)^2, (1 + i)^2, 0,0, \cdots)^T \qquad \textbf{(14)}$$

Similarly, all other possible spinors are obtained by changing the quantum number $m_J$. Nonzero components $B_i$ of the first excited state are proportional to $(v/c)^2$, and those of the fundamental state are proportional to $(v/c)$. This means that for velocities significantly lower than the speed of light, the components of the excited states become smaller with the increase in their order. Therefore, the excited states of the particle and of the antiparticle have physical meaning only in the limit $v \cong c$, i.e., under the conditions of high energy. Thus, the components of the excited states are zero only when the particle is in the fundamental state. When the velocity differs from zero, all excited states *become alive* with their components proportional to $(v/c)^n$ (*n* is the order of the excited state) and the particle is described by the wavevector given by superposition of all infinite states. Therefore, the solutions of the Majorana equation meet the principle of superposition of states in a natural way.

According to the theory of relativity, when the velocity of the particle approaches that of light, the inertial mass tends to infinity:

$$\lim_{v\to c} m = \lim_{v\to c} \gamma m_0 = \infty$$

However, in the Majorana theory, the mass of the excited states gradually decreases with the total angular quantum number:

$$m_{J_n} = \frac{m_0}{J_n + \frac{1}{2}} \quad \Rightarrow \quad m_0, \frac{m_0}{2}, \frac{m_0}{3}, \cdots$$

Hence, the relativistic form of the inertial mass becomes:

$$m_{J_n} = \frac{\gamma}{(n+1)} m_0 \quad n = 0,1,2,3, \cdots \qquad \textbf{(15)}$$

Equation **(15)** shows that the mass depends on the two seemingly unrelated variables: the particle velocity (by the Lorentz factor) and the excited state order. Considering these two variables, we can write the following limits:

$$\begin{cases} \lim\limits_{n\to\infty} m_{J_n} = 0 \\ \lim\limits_{v\to c} m_{J_n} = \infty \end{cases}$$

The order of the excited state depends on $(v/c)^n$, which can be considered as the statistical weight that the particle is described by the nth excited state. Hence, variables $\gamma$ and $n$ are mutually dependent. In principle, an upper limit for $n$ does not exist, and if speed $v$ is sufficiently close to the speed of light, states with high $n$ can be populated. Therefore, in the Majorana theory, the probability that the particle is in an excited state depends on its velocity. Referring to Eq. **(15)**, in the limit $v \to c$, the progressive increase of the Lorentz factor is mitigated by the simultaneous increase of order $n$ of the excited state. This result is topical since it is completely new for current relativistic quantum theories and suggests that, for high energy particles, a quantum transition from a bradyonic to a tachyonic behavior could be possible [21-25].

The probability that a particle is described by state $|\Psi_i\rangle$ is given by $c_i^2$, i.e., by the square of its linear combination coefficient:

$$|\varphi\rangle = c_1|\Psi_1\rangle + \cdots + c_i|\Psi_i\rangle + \cdots + c_n|\Psi_n\rangle$$

where $|\varphi\rangle$ is the eigenvector obtained by the linear combination of all possible states. The Majorana spinor can also be expressed as the linear combination of the fundamental state and all infinite excited states:

$$|\varphi\rangle^M = c_0|s\rangle + c_1|s+1\rangle + \cdots + c_n|s+n\rangle + \cdots \quad \textbf{(16)}$$

This is the reason why in the previous section we considered all components of the spinors as zero that are trivial solutions of the system of infinite linear differential equations. Therefore, the spinor in Eq. **(16)** is a vector with all nontrivial components, similar to the spinor in Eq. **(14)**. It is easy to verify that the coefficients of the linear combination in Eq. **(16)** are given by:

$$c_n = \sqrt{\left(\frac{v}{c}\right)^n - \left(\frac{v}{c}\right)^{n+1}} \quad \textbf{(17)}$$

Therefore, the probability that the nth state is occupied is $\left[\left(\frac{v}{c}\right)^n - \left(\frac{v}{c}\right)^{n+1}\right]$ (this formula is proved in the next section).

## 4 Majorana Spinors

From the solution of the Majorana equation, we get an infinite wavevector with infinite components, each concerning a given excited state. These solutions are represented by:

$$|J_n, m_{J_n}\rangle$$

where *n* is the order of the excited state, $J_n = s + n$, and $m_{J_n}$ is one of $(2J_n + 1)$ *z*-components of the total angular momentum. Then we can write the general state of the particle setting component $m_{J_n}$ as follows:

$$|j, 1/2\rangle = c_0|1/2,1/2\rangle + c_1|3/2,1/2\rangle + \cdots + c_n|(1/2+n), 1/2\rangle + \cdots \quad \textbf{(18)}$$

where $c_n = (0, \cdots, n, \cdots)$ are the coefficients of the linear combination. The explicit form of the ket in the linear combination in Eq. **(18)** is:

$$|1/2,1/2\rangle = (a, b, c, d, 0,0, \cdots)$$

$$|3/2,1/2\rangle = (0,0,0,0,a',b',c',d',e',f',g',h',0,0,\cdots)$$

$$|5/2,1/2\rangle = (0,0,0,0,0,0,0,0,0,0,0,0,a'',b'',c'',\cdots,n'',0,0,\cdots)$$

where the letters indicate the nontrivial components of the wavevectors. Obviously, the higher the quantum number $J_n$ is the greater the number of nontrivial components. Moreover, for general spinors $|j,1/2\rangle$ and $|j,-1/2\rangle$, all infinite wavevectors with infinite components are the nontrivial solutions of the Majorana equation (at least all four components are not zero). In compliance with the postulates of quantum mechanics, the square of the coefficients of the linear combination are the probabilities of occupation of each state belonging to the infinite sum in Eq. **(18)**. In the Majorana theory, this probability have to be proportional to $(v/c)^n$, and when $v \rightarrow c$ it increases progressively with order *n* of the excited state.

Using the definition of probability, we have:

$$p_n = \frac{(v/c)^n}{\sum_{n=0}^{\infty}(v/c)^n} \qquad \textbf{(19)}$$

The denominator is the usual geometric series, and since the ratio $v/c$ is lower than 1, it converges to $[1/(1-v/c)]$. Then, Eq. **(19)** can be written as:

$$p_n = \frac{(v/c)^n}{1/(1-v/c)} = \left[\left(\frac{v}{c}\right)^n - \left(\frac{v}{c}\right)^{n+1}\right] \qquad \textbf{(20)}$$

Equation **(20)** corresponds to the probabilistic coefficient that is used in Sect. **3**. If *n* is set, function $p_n$ has a maximum at well-determined $(v/c)^n$. Figure 1 shows the trend of this function for values *n=3* and *n=6*:

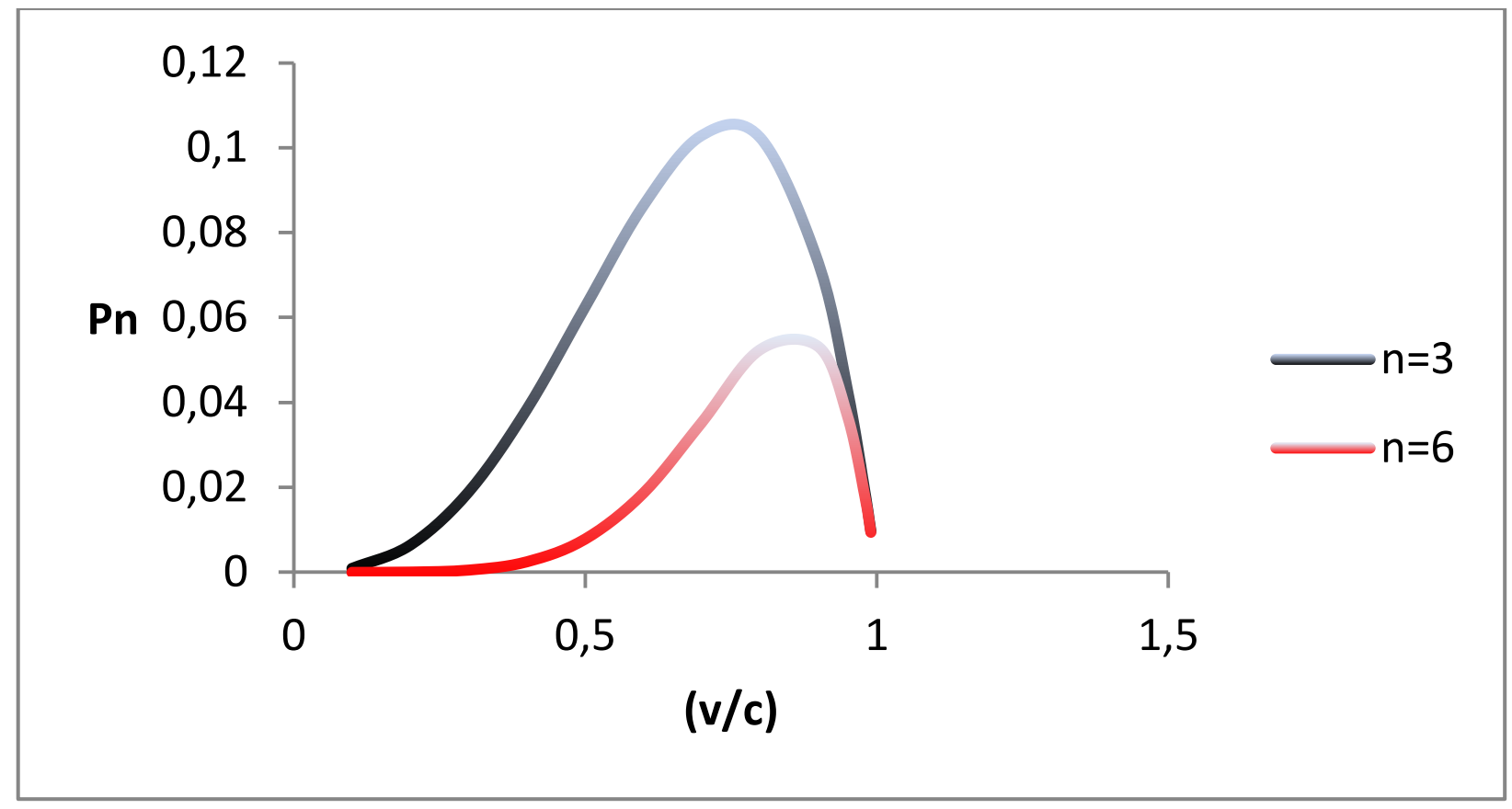

**Figure 1:** Occupation probability vs $(v/c)$

We can calculate velocity *v*, for which state *n* has the greatest probability to be occupied:

$$\frac{d}{dv}\left[\left(\frac{v}{c}\right)^{n}-\left(\frac{v}{c}\right)^{n+1}\right]=0$$

From this equation, we get:

$$(v/c)_{max}=\frac{n}{n+1} \qquad \textbf{(21)}$$

From Eq. **(21),** we obtain:

$$\lim_{n\to\infty}(v/c)_{max}=1 \quad \Rightarrow \quad v\to c$$

which proves that the probability function (Eq. **(20)**) is well defined and physically consistent with quantum theory developed so far. The general spinor in Eq. **(18)** can be rewritten as:

$$|j,1/2\rangle=\sqrt{1-(v/c)}|1/2,1/2\rangle+\cdots+\sqrt{\left[\left(\frac{v}{c}\right)^{n}-\left(\frac{v}{c}\right)^{n+1}\right]}|(1/2+n),1/2\rangle+\cdots$$

It is clear that values $(v/c)$ corresponding to the maximum probability tend to 1 when the excited state order increases. This increase occurs rapidly from *n=0* to *n=9*, and then becomes progressively slower. This means that when $v=0.90c$, the maximum probability of the excited states with $n>9$ remains almost constant. Let us study how the probability of state occupation changes when the particle velocity is set (i.e., changing *n*):

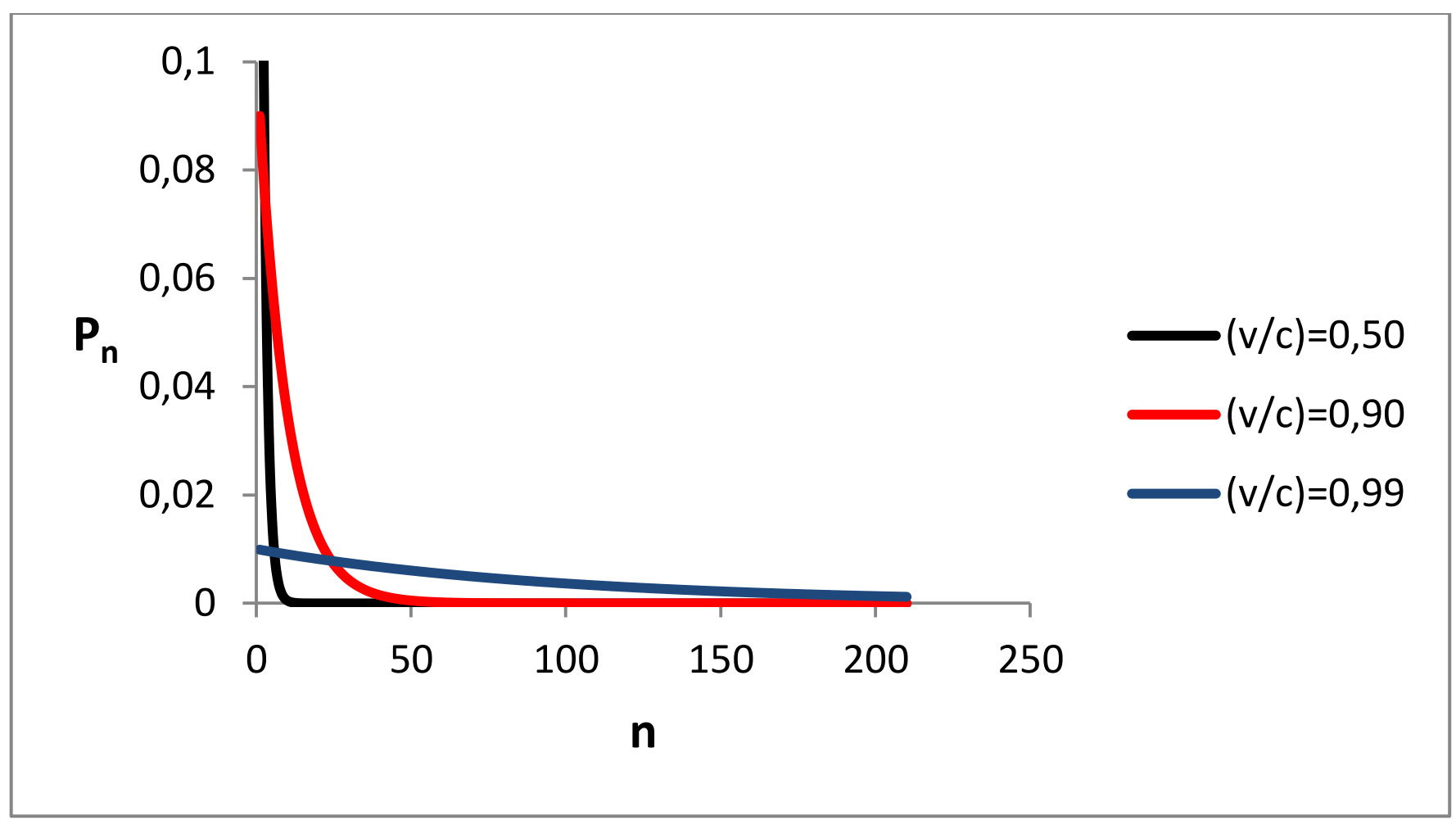

**Figure 2**: Occupation probability vs order of excited state

From Fig. 2, we see that with an increase in $(v/c)$, the probability that a state with low $n$ is occupied decreases rapidly. However, this decrease is homogeneously distributed over all excited states with high $n$. In the limit $v \to c$, all the excited states become equally probable and the *reality* of the particle is delocalized over all infinite states. In such a context, the uncertainty principle can *decide the fate* of the particle and keep it in a bradyonic state or bring it in a tachyonic state. Since order $n$ of the excited state is a positive integer, the trends shown in the three figures are discrete (even if they are presented using solid lines). Therefore, we can say that trend $p_n$, as a function of $n$, is the trend of a line spectrum, whose density increases with $n$.

## 5 Conclusion

In this study, an analytical method to solve the Majorana equation for a particle with half-integer spin has been proposed. The spinors of the fundamental and excited states have been explicitly calculated and commented with regard to their physical meaning. It has been also derived the formula which gives the probability of occupation of these states as a function of the particle velocity. The theory proves that for fermion particles traveling with a velocity very close to that of light (high energy particles) the Lorentz factor $\gamma$ does not tend to infinity and the transition from the bradyonic to the tachyonic state could be possible. The study of this transition will be done within the quantum field theory in the second step of this research, with the purpose to propose a new approach for a quantum theory of bradyons and tachyons [20]. The ultimate aim is to contribute to extending the current Standard Model in order to explain phenomena still lacking a scientific explanation, such as the flavor oscillation of neutrino and the possibility of its superluminal behavior [26-29], or the transitions through horizons of standard Black Holes [30-31].

## References

1. E. Majorana, Relativistic Theory of Particles with Arbitrary Intrinsic Angular Momentum, *Il Nuovo Cimento*, **9**, (1932) 335. – English Translation by C.A. Orzalesi in Technical Report, 792, University of Meryland (1968).
2. Ettore Majorana Scientific Papers, SIF, Bologna – Springer Berlin Heidelberg, New York (2006) – G.F. Bassani Editor (Council of Italian Physical Society) – Pisa.
3. D.M. Fradkin, Comment on a paper by Majorana concerning elementary particles, *Am. J. Phys*. **34**, (1966) 314.
4. I.M. Gelfand, A.M. Yaglom, General relativistic-invariant Equations and infinite-dimensional representations of the Lorentz group, Zh. Eksperim. i. Teor. Fiz., **18**, (1948) 707.
5. S. Esposito, The Physics of Ettore Majorana, Cambridge University Press (2014).
6. G. Bisiacchi, P. Budini, G. Calucci, Majorana Equations for Composite Systems, *Phys. Rev.*, **172** (5), (1967) 1508.
7. D. Tz. Stoyanov, Majorana Representations of the Lorentz Group and Infinite-Component Fields, *J. Math. Phys.*, **9**, (1968) 2146.
8. A. O. Barut, C. K. E. Schneider, R. Wilson. (1979) Quantum theory of infinite component fields, *J. Math. Phys.*, **20** (11), (1979) 2244.

9. A.O. Barut, I.H. Duru, Introduction of Internal Coordinates into the Infinite-Component Majorana Equation, *Proc. R. Soc. Lond. A*., **333**, (1973) 217.
10. X. Bekaert, M.R. de Traubenberg, M. Valenzuela, An infinite supermultiplet of massive higher-spin fields, arXiv 0904.2533v4 [hep-th] (2009).
11. P. A. Horváthy, M.S. Plyushchay, M. Valenzuela, Bosonized supersymmetry from the Majorana-Dirac-Staunton theory and massive higher-spin fields, *Phys. Rev. D*, **77**, 025017 (2008).
12. M.S. Plyushchay. Fractional spin. Majorana-Dirac field, *Phys. Lett. B*, **273** (3), (1991) 250.
13. M.S. Plyushchay, Relativistic particle with torsion, Majorana equation and fractional spin, *Phys. Lett. B*, **262**, (1991) 71.
14. E. C. G. Sudarshan, N. Mukunda, Quantum Theory of the Infinite-Component Majorana Field and the Relation of Spin and Statistics, *Phys. Rev. D*, **1** (2), (1970) 571.
15. X. Bekaert, S. Cnockaert, C. Iazeolla, M.A. Vasiliev, Nonlinear Higher Spin Theories in Various Domensions, arXiv hep-th/0503128 (2005).
16. L. Deleidi, M. Greselin, About Majorana's Unpublished Manuscripts on Relativistic Quantum Theory for Particles of Any Spin, Universal *J. Phys. Appl*., **9**, (2015) 168.
17. A. Bettini, Introduction to Elementary Particle Physics, Cambridge University Press, New York (2008).
18. P.A.M. Dirac, The Quantum Theory of the Electron, *Proc. R. Soc. A*, **117**, (1928) 610.
19. W. Pauli, Zur Quantenmechanik des magnetischen Elektrons, *Zeitschrift für Physik* **43**, (1927) 601.
20. D. McMahon, Quantum Field Theory, Mc Graw Hill (2008).
21. E. Recami, W.A. Rodrigues, Tachyons: may they have a role in elementary particle physics?, *Prog. In Particle and Nuclear Phys.*, **15**, (1985) 499.

22. E. Recami, Superluminal motion? A Bird's-eye view of the experimental situation, *Foundations of Physics*, **31** (7), (2001) 1119.

23. R.L. Dawe, K.C. Hines, The Physics of Tachyons, *Australian J. Phys.*, **45**, (1992) 591.

24. V. S. Olkhovsky, E. Recami, G. Salesi, Superluminal tunneling through two successive barriers, *Europhis. Lett.*, 57 (6), (2002) 879.

25. V.S. Olkhovsky, E. Recami, Recent developments in the time analysis of tunneling processes, *Phys. Reports*, **214**, (1992) 339.

26. D. Kirilova, Neutrinos from the Early Universe and Physics Beyond Standard Model, Open Phys., 13, (2016) 22.

27. E. Giannetto, G.D. Maccarone, R. Mignani, E. Recami, Are Muon Neutrinos faster-than-light particles?, *Phys. Lett. B*, **178** (1), (1986) 115.

28. R. Ehrlich, Faster-than-light speeds, tachyons, and the possibility of tachyonic neutrinos, *Am. J. Phys.*, **71**, (2003) 1109.

29. E. Recami, Classical Tachyons and Possible Applications, Rivista Nuovo Cimento, **9** (6), (1986) 1.

30. D.A. Kirzhnits, V.N. Sazonov, Faster-than-light motions and the special theory of relativity, Einstein collection, A75-43227 21-70, (1973) Moscow.

31. V. De Sabbata, M. Pavsic, E. Recami, A new view about black holes, *Lett. Nuovo Cim.*, **19**, (1977) 441.